\documentclass{article}

\usepackage{PRIMEarxiv}

\usepackage[utf8]{inputenc} % allow utf-8 input
\usepackage[T1]{fontenc}    % use 8-bit T1 fonts
\usepackage{hyperref}       % hyperlinks
\usepackage{url}            % simple URL typesetting
\usepackage{booktabs}       % professional-quality tables
\usepackage{amsfonts}       % blackboard math symbols
\usepackage{nicefrac}       % compact symbols for 1/2, etc.
\usepackage{microtype}      % microtypography
\usepackage{lipsum}
\usepackage{fancyhdr}       % header
\usepackage{graphicx}       % graphics
\usepackage{siunitx}
\usepackage{amsmath}
\usepackage{gensymb}
\usepackage{lineno}
\graphicspath{{media/}}     % organize your images and other figures under media/ folder

\title{Needle guidance with Doppler-tracked polarization-sensitive optical coherence tomography
}

\author{
  Danielle J. Harper \\
  Wellman Center for Photomedicine, Massachusetts General Hospital, 40 Blossom Street, Boston, MA 02114, USA \\
  Harvard Medical School, 25 Shattuck Street, Boston, MA 02115, USA \\
  \texttt{djharper@mgh.harvard.edu} \\
  \AND
  Yongjoo Kim \\
 Wellman Center for Photomedicine, Massachusetts General Hospital, 40 Blossom Street, Boston, MA 02114, USA \\
  Harvard Medical School, 25 Shattuck Street, Boston, MA 02115, USA \\
\texttt{ykim60@mgh.harvard.edu} \\
\And
 Alejandra Gómez-Ramírez \\
  Wellman Center for Photomedicine, Massachusetts General Hospital, 40 Blossom Street, Boston, MA 02114, USA \\
  School of Physics, Universidad Nacional de Colombia sede Medellín, Medellín, Colombia\\
 \texttt{alegomezram@unal.edu.co} \\
    \And
  Benjamin J. Vakoc \\
  Wellman Center for Photomedicine, Massachusetts General Hospital, 40 Blossom Street, Boston, MA 02114, USA \\
  Harvard Medical School, 25 Shattuck Street, Boston, MA 02115, USA\\Harvard-MIT Division of Health Sciences and Technology, 77 Massachusetts Avenue, Cambridge, MA 02139, USA \\
  \texttt{bvakoc@mgh.harvard.edu} \\
}
\begin{document}
\maketitle

\begin{abstract}

We demonstrate that a simple, unscanned polarization-sensitive optical coherence tomography needle probe can be used to perform layer identification in biological tissues. Broadband light from a laser centered at 1310 nm was sent through a fiber that was embedded into a needle, and analysis of the polarization state of the returning light after interference coupled with Doppler-based tracking allowed the calculation of phase retardation and optic axis orientation at each needle location. Proof-of-concept phase retardation mapping was shown in Atlantic salmon tissue, while axis orientation mapping was demonstrated in white shrimp tissue. The needle probe was then tested on the ex vivo porcine spine, where mock epidural procedures were performed. Our imaging results demonstrate that unscanned, Doppler-tracked polarization-sensitive optical coherence tomography imaging successfully identified the skin, subcutaneous tissue, and ligament layers, before successfully reaching the target of the epidural space. The addition of polarization-sensitive imaging into the bore of a needle probe therefore allows layer identification at deeper locations in the tissue.
\end{abstract}

\vspace{20pt}
\section{Introduction}\label{sec1}

While optical imaging in biological tissue is limited to a few millimeters \cite{finlayson2022depth}, needle probes can be used to access deeper sites. In these ``smart needles,'' the tip of the needle becomes the probe and feeds information back to the main system via an optical fiber. Needle-based Raman spectroscopy \cite{iping2015characterisation}, for example, has found applications in the evaluation of Duchenne muscular dystrophy  \cite{alix2022fiber} and the diagnosis of mitochondrial muscle disease \cite{alix2022application}. High-resolution photoacoustic endomicroscopy has shown cellular-level resolution images of mouse blood cells \cite{zhao2022ultrathin}. Needle-based confocal laser endomicroscopy is showing promise for the diagnosis and staging of lung cancer \cite{wijmans2019needle}. 

Preceding the development of each of these methods were demonstrations of needle probes based on optical coherence tomography (OCT) \cite{boppart1997forward,li2000imaging} developed for applications such as fine needle aspiration biopsy guidance \cite{goldberg2008automated} and adenocarcinoma identification \cite{swaan2019first}. However, despite its long history and the maturity of today's OCT technology, needle-based OCT has not been adopted in clinical practice. One of the primary reasons for this is the high cost of OCT instrumentation and needle-based OCT probes relative to the constraints of the candidate clinical applications. Miniaturization promises to reduce OCT system costs  \cite{rank2021miniaturizing}, creating a focused need for low-cost needle probes. To this end, scanned-beam probes based on mechanical actuators \cite{pan2001endoscopic,zara2004endoscopic} or fiber bundles \cite{wurster2018endoscopic} are less attractive than simple unscanned designs \cite{xie2013coronal}. However, unscanned OCT imaging data is difficult to interpret, especially in forward-viewing probes, resulting in a performance penalty that may be prohibitive.   

When a scanned-beam OCT system requires additional imaging contrast, polarization-sensitive (PS) methods \cite{hee1992polarization} often provide a solution. PS-OCT reveals tissue birefringence features that are invisible to structural (non-PS) OCT imaging. PS-OCT has been used to image skeletal muscle \cite{mcbride2022vivo}, heart \cite{zhao2018integrated,zhao2021polarization} and brain \cite{depaoli2022endoscopic}. In all of these procedures, PS-OCT was added to a scanned-beam approach. To our knowledge, PS-OCT has not been studied for use in unscanned probes.  

In this work, we demonstrate a simple, unscanned PS-OCT needle probe that could be used as a tissue layer identification tool in layered samples. To provide an exemplary application, we explore epidural space identification during nerve block administration, an application in which scanned-beam OCT \cite{kuo2015fiber,kuo2017vivo,wang2022epidural} and PS-OCT \cite{ding2016imaging} methods were shown to be promising. We additionally demonstrate the method more generally by imaging in tissues selected for  layered retardance (salmon) and layered optic axis (shrimp) archetypes. Included in this work is a means for using Doppler-based position-tracking  to transform the PS-OCT A-line data into a spatial visualization that mimics a cross-sectional view through the plane of the needle, aiding in interpretation. These results suggest that unscanned PS-OCT data can overcome the ambiguity of unscanned structural OCT data, opening the door to using low-cost probes in cost-sensitive applications. 

\section{Materials and Methods}
\subsection{OCT system}
We used a 1310-nm PS-OCT system as described by Vakoc et al. \cite{vakoc2005phase,vakoc2009three}. The swept-source laser was built in-house and was based on a polygon scanning mirror. The laser sweep rate was 50 \si{\kilo\hertz}. An acousto-optic frequency shifter in the reference arm provided an acquisition-limited scan range of 5.8 mm \cite{yun2004removing}. Owing to the need to visualize the needle tip itself in the image, the scan range allocated to the tissue was 4.7 mm. The axial resolution of the system was measured to be 10.6 \si{\micro\meter} in air. An electro-optic modulator was used to alternate the polarization states of light incident on the sample between successive A-lines. 

\subsection{PS-OCT imaging needles}

Prototypes of single-use imaging needles were constructed by embedding SMF-28 fiber within needle bores using heat-cured epoxy. Fibers with a 900-\si{\micro\meter}-diameter jacket were stripped at one end down to their 245-\si{\micro\meter}-diameter buffer layer and the stripped end was fed through the needle until the tip passed through the bevel. The needle hub was then filled with the epoxy and connected to the barrel. Using a plunger, the epoxy was pushed through the lumen until it was visible at the tip of the shaft. Excess epoxy at the tip was removed. The fiber was then cleaved and pulled back to a position where it was slightly below the level of the tip, but still visible within the bevel. The assembly was then tightly secured and placed in an incubator to cure at 80 \degree C. This process was used to construct 29 gauge, 1.2 cm long and 22 gauge, 6.5 cm long needles. 

After curing, the fibers were spliced to FC/APC connectors, allowing optical connection to the PS-OCT system. The total optical path length of the sample arm, including each probe, was matched to the optical path length of the reference arm.   

\subsection{Biological sample measurement}

\begin{figure}[tb]
\begin{center}
\begin{tabular}{c}
\includegraphics[width=16.1 cm]{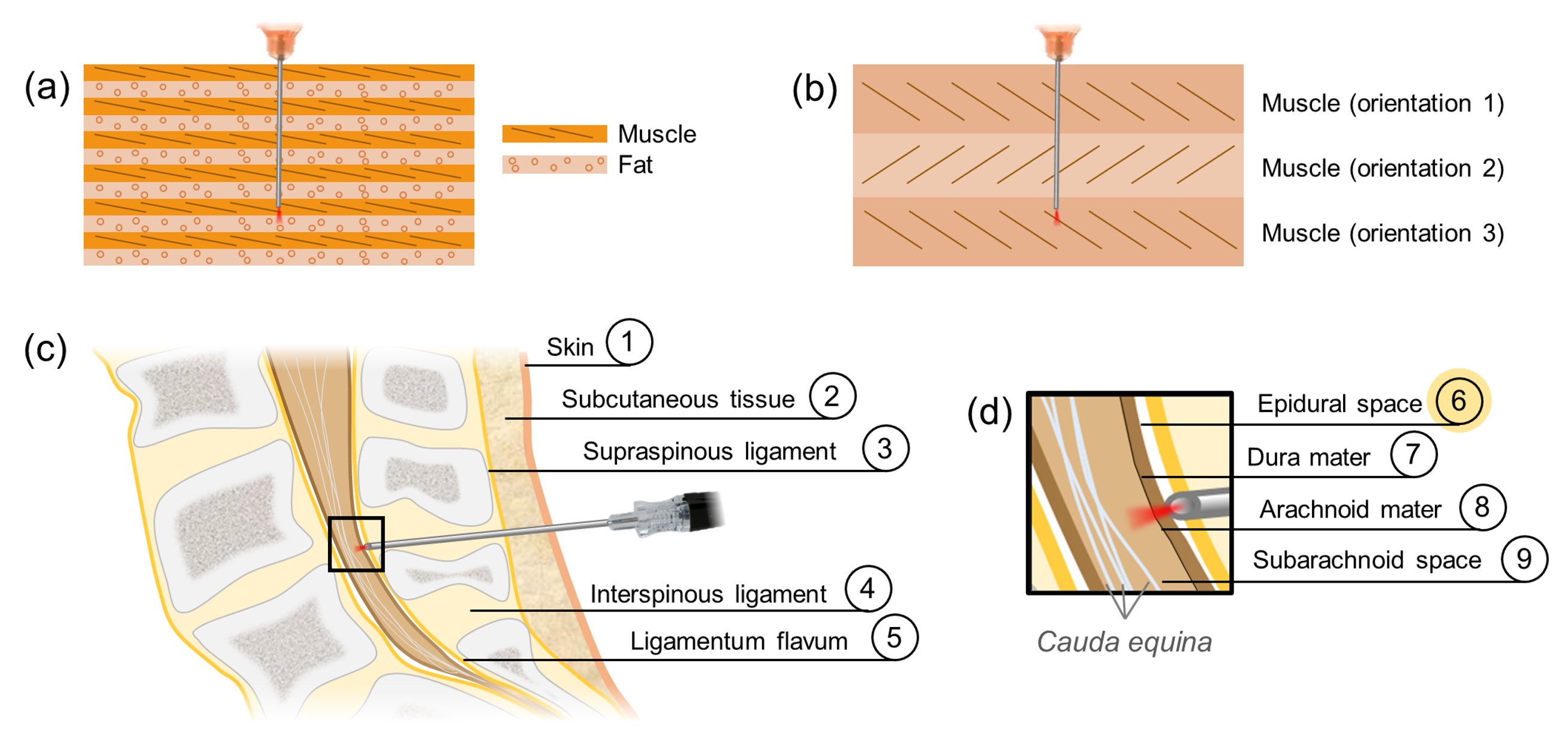}
\end{tabular}
\end{center}
\caption 
{ \label{fig1} Biological samples imaged in this work. a-b) Depiction of the sources of polarization-sensitive optical coherence tomography contrast in salmon (a) and shrimp (b) tissues. In salmon, highly oriented muscle induces large phase retardation of the incident light. In contrast, weakly oriented fat induces very little phase retardation. In shrimp, each muscle layer induces similarly high phase retardation, but the orientation of the fibers in each layer is distinct. c-d) Path of a needle during the epidural procedure. The needle must traverse the skin, subcutaneous tissue (fat), and three ligament layers (c) before reaching (d) the epidural space (layer 6). Care must be taken not to puncture the dura mater and the arachnoid mater, as reaching the subarachnoid space can result in a cerebral spinal fluid leak.} 
\end{figure} 

\subsubsection{Atlantic salmon for phase retardation measurement}

We first demonstrated the ability of needle-based PS-OCT to perform layer identification in tissue based on phase retardation contrast. As a sample, we selected Atlantic salmon muscle, which comprises alternating sheets of highly unidirectionally-oriented myomeres and relatively unoriented fatty connective tissue [see Fig. \ref{fig1}(a)] \cite{listrat2016muscle}. Retardance of the phase of one polarization state with respect to the other is induced within the myomeric layers, but not in the connective tissue. A 29 gauge, 1.2 cm long probe was pushed through these alternating tissue layers (five muscle and four connective tissue) over a duration of 6.5 \si{s} before being retracted in the same amount of time. PS-OCT imaging data were collected throughout. 

\subsubsection{White shrimp for optic axis orientation measurement}

To demonstrate contrast based on optic axis orientation, we selected white shrimp muscle tissue as a sample. While all shrimp muscle tissue is birefringent, the orientation of the muscle fibers is not unidirectional. Muscle layers have different fiber orientations, with discrete boundaries between them \cite{zhang2019genome}. An illustration of this can be found in Fig. \ref{fig1}(b).

The shell was peeled from the white shrimp body to expose the muscle below. A 29 gauge, 1.2 cm long probe was pushed through three muscle layers, each with its own muscle fiber orientation, and was subsequently retracted. The shrimp tissue was imaged for a total of 13 seconds.

\subsubsection{Porcine lower lumbar spine for mock epidural procedure}

The lower lumbar regions of three swine were harvested immediately post-mortem. Full needle insertions of a 22 gauge, 6.5 cm long needle probe were performed through the skin, subcutaneous tissue, and the ligament layers (Fig. \ref{fig1} (c)) to reach the epidural space as highlighted in Fig. \ref{fig1} (d). Successful epidural space identification was characterized by the co-localization of the loss of optic axis information with a spike in the intensity reading at that location. The intensity signal spike results from the mismatch in refractive index between the tissue-coated needle tip and the empty epidural space and also from the additional fat content in the epidural space in the pig \cite{pleticha2013pig}. Data was acquired in the lower lumbar for a total of 19 seconds.

\subsection{PS-OCT image processing}

PS-OCT image processing was performed to obtain intensity, phase retardation, and optic axis profiles for each A-line. Spectral binning (with five bins) was used to reduce the effects of polarization mode dispersion \cite{villiger2013spectral}. The intermediate output after standard PS-OCT image processing was a series of A-lines. The A-lines were concatenated into a single image giving an M-mode-like image representation, with time on the x-axis and depth on the y-axis. This representation was termed "longitudinal scans" by Xie et al. \cite{xie2013coronal}, and we will refer to this method of image visualization as a "needle-referenced map" in this work.

\subsection{Doppler tracking}

The phase information obtained for each polarization state was stored, and the difference between each subsequent A-line was computed for one of the polarization states. Any phase difference measured to be more than 2.5 rad was classified as phase jitter from the laser and was therefore set to zero. At each depth location, phase difference averaging was conducted via a sliding 72 A-line median filter, corresponding to a time average of 1.4 ms. As the distance from the zero delay to the reconstructed signal originating from the needle tip itself was constant, the observed phase difference measured at the needle tip location was subtracted from the phase difference at each depth below, correcting for slow phase drift. As each tissue layer was considered to be moving only relative to the needle tip and not relative to other tissue layers, a further average (mean) was calculated over 46 pixels below the surface of the needle tip, corresponding to a depth of 86 \si{\micro\meter}. With a single phase difference value between each A-line, $\Delta \varphi$, the relative axial displacement of the needle between two A-lines, $d(z)$, could then be measured as 

\begin{equation}
d(z)=\lambda_c \Delta \varphi/(4 \pi n) \cite{leitgeb2014doppler}
\end{equation}

\noindent where $\lambda_c$ is the central wavelength of the OCT light source, and the refractive index, $n$, was assumed to be 1.37 \cite{bashkatov2011optical}.

\subsection{Data visualization}

Needle-referenced maps were converted into surface-referenced maps with the aid of the Doppler-tracked position. In the needle-referenced map (x-z image), the z-axis represents the axis of the output light beam, as in traditional OCT, while the x-axis represents both time and depth, and this depth is relative to the needle tip. To reunite the two "depth" parameters into a single axis, we transform from the reference frame of the needle to the reference frame of the tissue surface using the Doppler-tracked distances. In the surface-referenced map, which we show in video format, the z-axis updates as the needle moves with respect to the tissue, and the x-axis updates constantly with time. The gradient of any resulting lines that appear due to the contrast between two boundaries, therefore, indicates the velocity of the needle, and the apparent "thickness" of the tissue indicates the time spent at that position, which can be used as a measurement of the confidence of the reading. The x-axis was mirrored around the needle tip to provide a more visually intuitive representation of the data, and the x-axis was no longer updated after the needle tip passed that layer of the tissue.

\section{Results} 
\subsection{Retardance mapping}

\begin{figure}
\begin{center}
\begin{tabular}{c}
\includegraphics[height=10.0 cm]{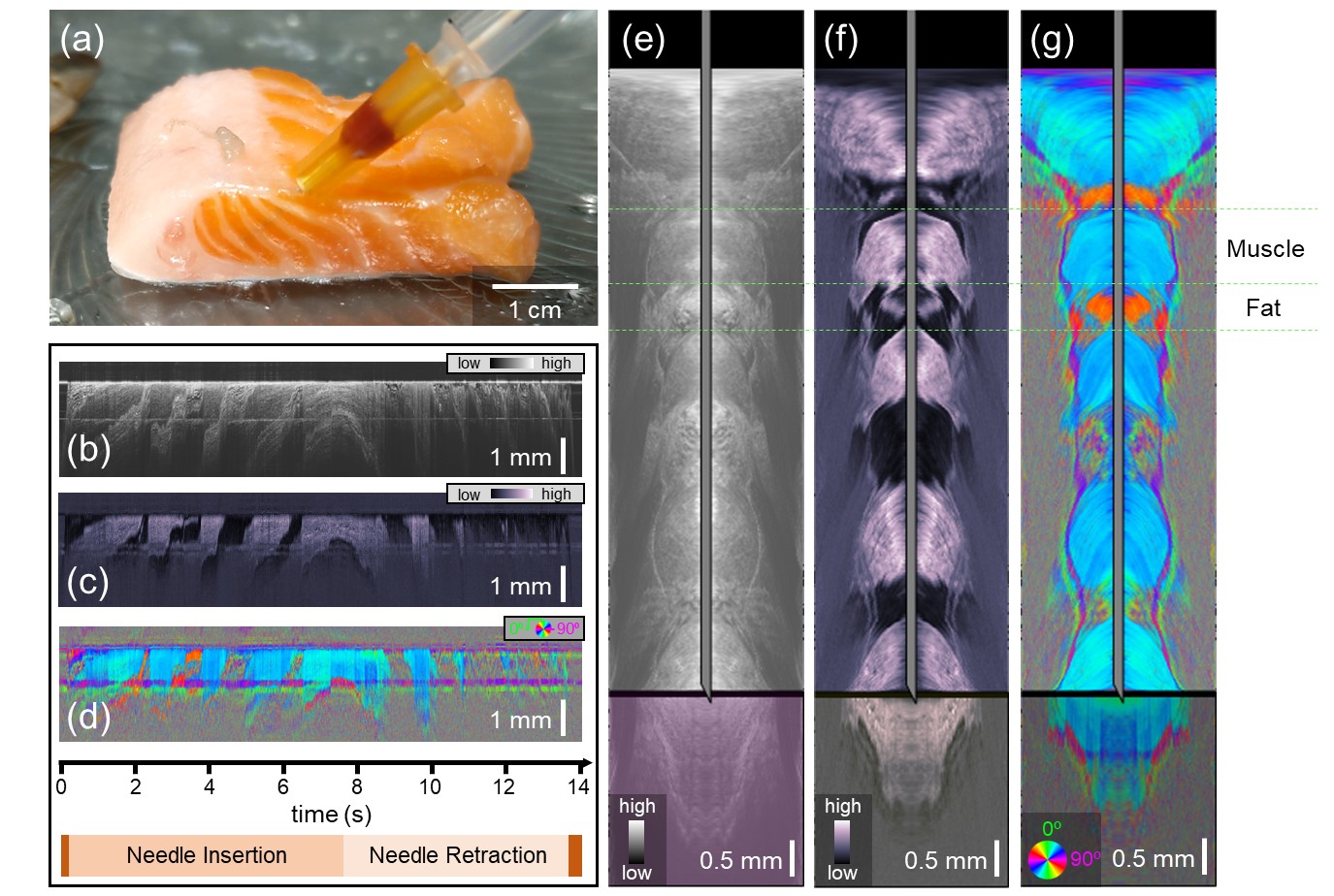}
\end{tabular}
\end{center}
\caption 
{ \label{fig2} Polarization-sensitive optical coherence tomography-based reconstructions of salmon tissue. a) Photo of the needle probe at its full insertion point in the salmon tissue. b-d) Needle-referenced map indicating the backscattered intensity (b), phase retardation (c), and optic axis orientation (d) of the salmon tissue. Time axis and needle insertion/retraction indicators apply to (b-d). e-g) Still frames from Video 1. Surface-referenced maps of intensity (e), phase retardation (f), and optic axis orientation (g). The transformation from the needle reference frame to the surface reference frame was made possible by Doppler tracking.} 
\end{figure}

The salmon provided layers of birefringent (muscle) and non-birefringent (fat) tissues, and was used to demonstrate retardance-based contrast (Fig. \ref{fig2}). Figure \ref{fig2}(a) shows a photograph when the needle was at maximum insertion. It can be seen from the side view that the needle passed through a total of five highly oriented muscle layers (pink/orange) and four connective tissue layers (off-white). The PS-OCT imaging data acquired from the probe as it was inserted into the tissue provided needle-referenced maps based on intensity (Fig. \ref{fig2}(b)), phase retardation (Fig. \ref{fig2}(c)) and optic axis orientation (Fig. \ref{fig2}(d)). In this sample, there is an intensity-based contrast between the layers, but the most striking contrast between the alternating layers comes from the phase retardation time series (Fig. \ref{fig2}(c)). The optic axis data (Fig. \ref{fig2}(d)) shows a similar orientation in each muscle layer. Optic axis data associated with fat is noisy, as expected. Using Doppler tracking, the needle-referenced maps in Fig. \ref{fig2}(b-d) were converted into surface-referenced maps, where the location of the needle tip with reference to the tissue surface is known. Snapshots of the intensity, phase retardation, and axis orientation at the location of maximum needle insertion can be found in Fig. \ref{fig2}(e), (f), and (g), respectively. Real-time video data showing the full insertion and pullback of the needle can be found in Video 1.

\subsection{Optic axis orientation mapping}
\begin{figure}
\begin{center}
\begin{tabular}{c}
\includegraphics[height=10.0 cm]{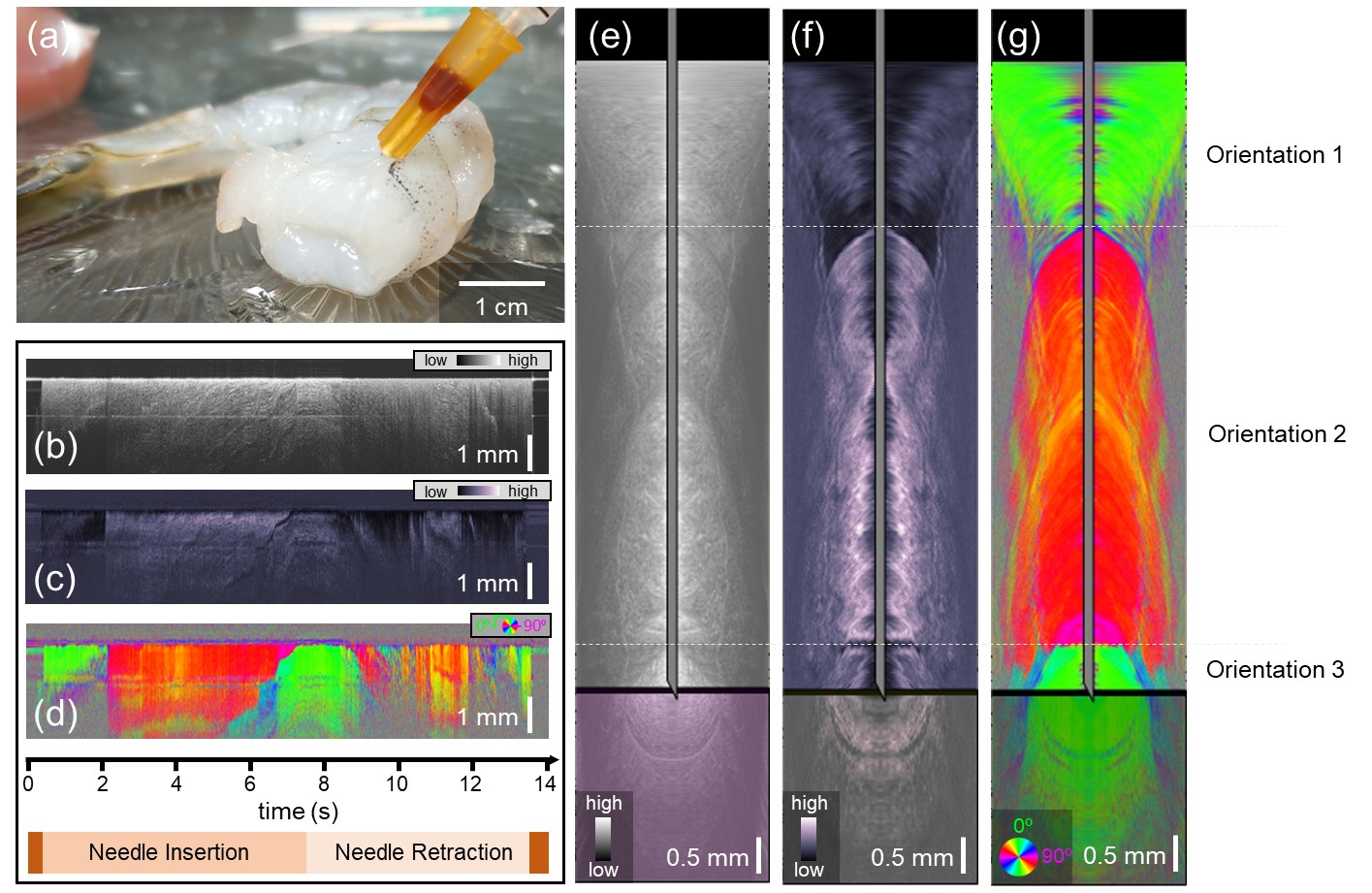}
\end{tabular}
\end{center}
\caption 
{ \label{fig3}
Polarization-sensitive optical coherence tomography-based reconstructions of shrimp tissue. a) Photo of the needle probe at its full insertion point in the shrimp tissue. b-d) Needle-referenced map indicating the backscattered intensity (b), phase retardation (c), and optic axis orientation (d) of the shrimp tissue. Time axis and needle insertion/retraction indicators apply to (b-d). e-g) Still frames from Video 2. Surface-referenced maps of intensity (e), phase retardation (f), and optic axis orientation (g). The transformation from the needle reference frame to the surface reference frame was made possible by Doppler tracking.} 
\end{figure}

The white shrimp provided layers of birefringent muscle tissues with different fiber orientations, and was used to demonstrate optic-axis-based contrast (Fig. \ref{fig3}). Figure \ref{fig3}(a) shows a photograph of when the needle was at its maximum insertion. The needle passed through three layers of muscle. The needle-referenced maps based on intensity (Fig. \ref{fig3}(b)), phase retardation (Fig. \ref{fig3}(c)) and optic axis orientation (Fig. \ref{fig3}(d)) show that, in this sample, the most striking contrast between the alternating layers comes from the axis orientation time series (Fig. \ref{fig3}(d)). Since each layer comprises muscle, both the intensity and phase retardation data remain relatively high in each layer. The optic axis data (Fig. \ref{fig3}(d)) clearly allows differentiation between the distinct layers of the shrimp muscle. Using Doppler tracking, the needle-reference maps in Fig. \ref{fig3}(b-d) were converted into surface-referenced maps, where the location of the needle tip with reference to the tissue surface is known. Snapshots of the intensity, phase retardation, and axis orientation at the location of maximum needle insertion can be found in Fig. \ref{fig3}(e), (f) and (g), respectively. Real-time video data showing the full insertion and pullback of the needle can be found in Video 2. 

\subsection{Proof-of-concept in the lumbar spine}

\begin{figure}
\begin{center}
\begin{tabular}{c}
\includegraphics[height=10.0 cm]{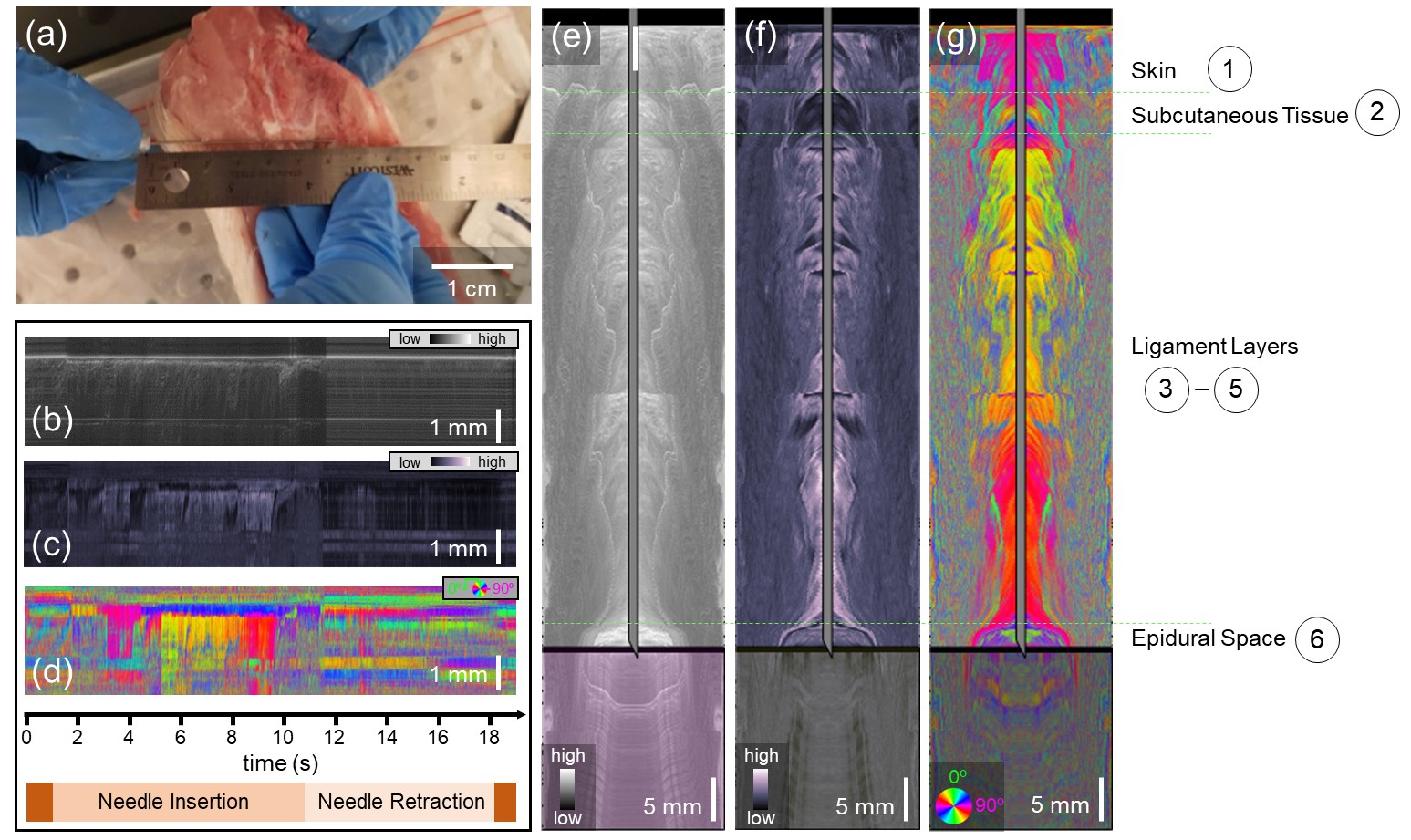}
\end{tabular}
\end{center}
\caption 
{ \label{fig4}
Polarization-sensitive optical coherence tomography-based reconstructions of porcine lower lumbar spine. a) Photo of the needle probe on top of the lower lumbar sample. The insertion was performed through the tissue. b-d) Needle-referenced map indicating the backscattered intensity (b), phase retardation (c), and optic axis orientation (d) of the spine tissue. Time axis and needle insertion/retraction indicators apply to (b-d). e-g) Still frames from Video 3.  Surface-referenced maps of intensity (e), phase retardation (f), and optic axis orientation (g). The transformation from the needle reference frame to the surface reference frame was made possible by Doppler tracking.} 
\end{figure}

The results of needle-based PS-OCT in the lower lumbar can be found Fig. \ref{fig4}. Figure \ref{fig4}(a) indicated the location of maximum insertion of the needle probe adjacent to the lower lumbar tissue specimen, although the mock epidural procedure was performed through the tissue. The needle was inserted over 6 cm into the tissue and subsequently removed in 19 seconds, resulting in a higher needle velocity than in the salmon and shrimp tissues. The needle-referenced maps based on intensity (Fig. \ref{fig4}(b)), phase retardation (Fig. \ref{fig4}(c)), and optic axis orientation (Fig. \ref{fig4}(d)) show a contrast between the subsequent layers, but layer identification remains challenging prior to conversion to the surface-referenced map. Snapshots of the intensity, phase retardation, and axis orientation at the location of maximum needle insertion can be found in Fig. \ref{fig4}(e), (f) and (g), respectively. Real-time video data showing the full insertion and pullback of the needle can be found in Video 3. The combination of these three datasets allows us to draw conclusions regarding the appearance of each of the porcine spine tissue layers when observed with PS-OCT. The needle first penetrates the skin, characterized by relatively low phase retardation and defined optic axis orientation. The subcutaneous tissue, without a defined structural organization, is characterized by a drop to very low phase retardation and corresponding weakly defined (non-constant) optic axis orientation. The three ligament layers all induce greater phase retardation and have a very defined optic axis orientation. A mild drift is observed in the optic axis orientation, most likely due to the slight bending of the fiber as the needle penetrates the tissue. Finally, the target of the epidural space is clearly characterized by the simultaneous increase in intensity coupled with a loss of signal in the phase retardation and optic axis orientation images. The increase in intensity in the epidural space is an artifact of the refractive index mismatch between the now-dirty fiber tip and the surrounding space. The location of the epidural space identified by PS-OCT occurred at the location where a loss of resistance was felt when pushing the needle through the tissue, resulting in the knowledge that the correct location had been identified.

\section{Discussion} 

This study asked if a simple, unscanned PS-OCT imaging needle probe alone was capable of providing a reliable report of the anatomical location of the needle's tip. Based on induced phase retardance mapping, the first PS-OCT probe successfully differentiated between the muscle and fat layers in salmon. In the shrimp muscle, the discrete boundaries between differently oriented muscle layers were clearly discernible using the axis orientation data. In the mock nerve block administration procedure, the contrast between the highly retarding, strongly oriented ligament and the orderless epidural space provided an ideal signal on which a user could determine when the needle enters the epidural space. PS-OCT could clearly identify this boundary. Doppler-based data analysis allowed the absolute position of the needle with reference to the tissue surface to be tracked in both the forward and backward direction (insertion and retraction of the needle).

We chose to remove all focusing and scanning optics to allow us to make a very cheap, disposable probe, increasing the likelihood of eventual clinical and research adoption. The absence of the focusing optics led to a diverging beam exiting the probe. At first look, it may seem as though this significantly reduces the signal-to-noise ratio (SNR). However, as the single-mode fiber has a similar numerical aperture to the lenses commonly used in OCT (0.14), we did not observe a prohibitive level of signal degradation over the penetration depth of the tissue. In a traditional OCT system, the design of the relay and focusing optics gives control over the depth of focus and the spot size, i.e., the lateral resolution as a function of depth. In the needle probe, such parameters are dictated by the choice of optical fiber. Should higher lateral resolution be desired, a single-mode fiber with a smaller core diameter could be used in place of the SMF-28 used in this work. Traditional OCT microscopes also allow the beam waist to be offset from the final optic, making use of the full confocal parameter. In contrast, the needle probe has the beam waist equivalent situated right at the needle tip, limiting the depth of focus to only the Rayleigh range, or half the confocal parameter. As the needle probe scans in depth, however, this does not present a challenge for imaging. As the needle advances further into the tissue, eventually every depth will be imaged at the beam waist (up to the point of maximum insertion). 

The signal-to-noise ratio (SNR) of the images given by the needle probe is higher upon insertion than it is during retraction, and the effect of this SNR degradation is dependent upon the needle diameter. When using the smaller (29 gauge) needle, the hole created by the needle itself was able to close naturally upon needle retraction. This allowed robust signals to be collected while the needle was retracted. This was not the case with the 22 gauge needle. Penetration of the tissue with this larger needle left a hole in the tissue such that the forward-looking needle probe was no longer in contact with the tissue and therefore led to weakened OCT signals upon retraction. This is demonstrated in Fig. \ref{fig4}(c-e). The Doppler-derived needle tip location data was less affected by this due to the large smoothing and averaging operations performed on the data. A potential solution would be to modify the probe design to create a side-looking beam made possible by total internal reflection. While this would again increase the probe complexity, it may be necessary to pursue in vivo imaging during the epidural nerve block administration procedure if it is likely that the needle will be repeatedly moved inwards and outwards. 

We chose to demonstrate this technique in the context of epidural guidance because it allows comparison with a large body of literature spanning a variety of approaches in addition to the aforementioned OCT-based methods \cite{ding2016imaging, kuo2015fiber,kuo2017vivo,wang2022epidural}. For example, an ultrasound probe was developed for this purpose, monitoring the distance to the spinous processes \cite{zhang2018toward}, but resolution limits mean that ultrasound cannot provide layer-specific identification. To reach the desired resolution, attention has turned to optical methods. A simple probe based that illuminates the tissue with blue light and measures the intensity of the reflected light successfully identified the epidural space \cite{lin2019optically}, but cannot give advanced warning as the boundary is approached. Mimicking the current approach, Carotenuto et al. proposed an optical measurement of the force applied to the tip of the needle as it passes through each layer using a fiber-Bragg grating \cite{carotenuto2018smart,carotenuto2019optical} and work is currently ongoing to create a biocompatible version of this probe. The plethora of guidance tools currently under development may help to facilitate nerve block administration in emergency settings, lowering the level of experience needed to perform the procedure successfully \cite{thangamuthu2013epidural,ismail2021failure}.  

We believe that needle-based PS-OCT has a high potential to be miniaturized and could be of value also in scenarios where the form factor must be small. As a step towards this goal, we have demonstrated that a simple probe without any scanning or focusing optics is sufficient for layer identification. However, this probe is still attached to a benchtop OCT imaging system. As a guidance tool for needle insertion, the accuracy of depth location does not need to be the full 10 \si{\micro\meter} that a system such as this offers, and the 50 kHz repetition rate is also much faster than necessary. Instead, in the interest of reducing the OCT system's physical and financial footprints, it may be of value to change the light source to a simple superluminescent diode with only several nanometers of optical bandwidth. We believe that a simple OCT system based on such a light source would be sufficient for this application and would even provide a signal-to-noise ratio benefit by reducing the optical bandwidth \cite{harper2022relationship}.

In addition to system simplification, it is also important to consider how best to maximize the number of different imaging contrast modalities that could be extracted from the same needle probe imaging data using post-processing techniques. Attenuation coefficient mapping \cite{kut2015detection} may be particularly useful in applications where the main goal is to differentiate between muscle and fatty tissue. In the intensity needle-referenced map scan in Fig. \ref{fig2}(c), it can be observed that the connective tissue layers can be observed below the muscle layers. Still, the muscle layers cannot be observed below the fat layers of a similar thickness. This is indicative that the attenuation coefficient of the fat is higher than that of the muscle and could serve as a useful parameter for future identification. Spectroscopic \cite{morgner2000spectroscopic} or hyperspectral \cite{harper2019hyperspectral} OCT post-processing could also be applied if the sample was interrogated with an optical bandwidth over which the backscattering coefficient changes dependent upon the tissue type. This work used a light source centered at 1310 nm, which is not the best wavelength for spectroscopic imaging. Future work targeting vasculature may wish to consider the reduction of wavelength as low as the visible light range, which would allow oximetry to be performed \cite{yi2013visible}. 

As more contrast modes are added to the data output, further consideration will need to be given to ensure succinct visualization for the end-user. While it would be possible to obtain depth information in an analog fashion, i.e., with direct marking on the needle, the Doppler tracking allows us to digitize this to create more interpretable visualizations. The chosen data visualization method for this study was to place absolute depth on the z-axis and time on the x-axis for each intensity, phase retardation, and optic axis orientation signals. Inherently encoded within the x-axis here is a type of confidence interval; the longer the needle acquires data at a particular location in depth, the thicker the signal appears along the x-axis. Care should be taken not to interpret this as two-dimensional imaging data, as this is inherently a one-dimensional imaging technique. To display only the phase retardance and optic axis data as a function of absolute depth, only a 2D plot is required. Our rationale for including the additional dimension was to make the imaging data more interpretable by the human eye and to include this idea of a confidence interval within the visual representation. It may be that other visualization methods, perhaps enhanced by machine learning or other image classification tools, are more appropriate for other applications. 

\section{Conclusion} 
In this work, we have demonstrated that a simple, forward-looking, unscanned PS-OCT fiber probe is sufficient to track the anatomical location of a needle tip within tissue. After demonstrating successful retardance and optic axis mapping in salmon and shrimp, the needle probe was used to identify the epidural space in the ex vivo swine. This makes a promising case for reducing the complexity of PS-OCT systems without compromising the utility of the information they generate in needle-based applications.

\subsection*{Acknowledgments}

The authors would like to thank Abigail Gregg, Mark Randolph and Fernando Guastaldi for assistance with spine tissue collection and preparation, and Ahhyun Stephanie Nam and Mohsen Erfansadeh for assistance with the post-processing framework. Funding from U.S. Department of Defense through the Military Medical Photonics Program (FA9550-20-1-0063) and the National Institute of Health (P41 EB015903) is gratefully acknowledged. Danielle J. Harper would like to acknowledge salary support from the Optica Thomas F. Deutsch Fellowship.

\subsection*{Conflict of interest}

The authors have no potential conflicts of interest to disclose.

\subsection*{Data availability}

Data is available upon request.

%\bibliography{WileyNJD-AMA}%

%Bibliography
\bibliographystyle{unsrt}  
\bibliography{sample}

\end{document}